\documentclass[a4paper]{aa}
\usepackage{graphicx}
\usepackage{txfonts}

\usepackage[english]{babel}

\usepackage{natbib}
\bibpunct{(}{)}{;}{a}{}{,}

\begin{document}

\title{Stirring up the dust: A dynamical model for halo-like dust clouds in transitional disks}

\titlerunning{A dynamical model for halo-like dust clouds in transitional disks}

\author{S. Krijt\inst{1,2} \and C. Dominik\inst{2,3}}

\institute{Leiden Observatory, Niels Bohrweg 2, 2333 CA Leiden, The Netherlands\\
              \email{krijt@strw.leidenuniv.nl}\label{inst1}
\and Astronomical Institute ``Anton Pannekoek'', University of Amsterdam, PO Box 94249, 1090 GE Amsterdam, The Netherlands\label{inst2}
\and Department of Astrophysics, Radboud University Nijmegen, PO Box 9010, 6500 GL Nijmegen, The Netherlands\label{inst3}
}

\date{Received 21 Februari 2011 / Accepted 1 May 2011}
\abstract
{A small number of young stellar objects show signs of a halo-like structure of optically thin dust, in addition to a circumstellar disk. This halo or torus is located within a few AU of the star, but its origin has not yet been understood.}
{A dynamically excited cloud of planetesimals colliding to eventually form dust could produce such a structure. The cause of the dynamical excitation could be one or more planets, perhaps on eccentric orbits, or a migrating planet. This work investigates an inwardly migrating planet that is dynamically scattering planetesimals as a possible cause for the observed structures. If this mechanism is responsible, the observed halo-like structure could be used to infer the existence of planets in these systems.}
{We present analytical estimates on the maximum inclination reached owing to dynamical interactions between planetesimals and a migrating planet. In addition, a symplectic integrator is used to simulate the effect of a migrating planet on a population of planetesimals. Collision time scales are estimated for the resulting population of planetesimals and the size distribution of the dust created in catastrophic collisions is determined.}
{It is found that an inwardly migrating planet is only able to scatter the material it encounters to highly-inclined orbits if the material is on an eccentric orbit. Such eccentric orbits can be the result of resonance trapping and eccentricity pumping. Simulations show that for a certain range of migration rates and planet masses, resonance capture combined with planetary migration indeed causes the planetesimals to reach eccentric orbits and subsequently get scattered to highly-inclined orbits. The size distribution of the resulting dust is calculated determined to find the total mass and optical depth, which are found to compare reasonably well with the observed structures.}
{Dynamical scattering of planetesimals caused by a planet migrating in, followed by the grinding down of these planetesimals to dust grains, appears to be a promising explanation for the inferred circumstellar dust clouds. Further study is needed to see if the haloes can be used to infer the presence of planets.} 


\maketitle

\section{Introduction}
The vast majority of stars are born with a circumstellar disk. Observations show that in young clusters the percentage of stars with disks is close to 100\%. The typical life times for disks appear to be around 6 Myr \citep{haisch01}. Circumstellar disks around young stars are believed to be the sites of planet formation. Solids in the disk grow to form larger objects, planetesimals, and eventually planets. Eventually, after the gas disappears, a planetary system will remain, containing planets and, if there is room in the dynamical structure, a debris belt.

Observations of protoplanetary disks at ever higher spatial resolution have provided promising indirect indications of ongoing planet formation \citep[e.g.][]{thalmann10,jang-condell10}. Another strong indication of the presence of planets comes from structure observed in the dust that is still visible at later times in debris disks \citep{wilner02,wyatt03,wyatt08}.

However, since most of the disks are difficult to resolve, studies also focus on understanding the system's spectral energy distribution (SED). Comparing the shape of the SED to radiative transfer models gives information on the geometry of the disk and on the temperatures and particles found in the different parts of the circumstellar disk \citep{dullemond07}.

Over the past two decades, studying SEDs has improved our knowledge of circumstellar disks significantly and has made us aware of structures as flaring disks, puffed up inner rims, gaps, and holes \citep[e.g.][]{dullemond01,dullemond07}. Holes and gaps are considered to be possible signposts of planet formation \citep{armitage10}.

Recent studies of (pre-)transitional disks indicate a possible new structure. Several young stellar objects (YSOs) appear to show signs of an optically thin population of dust with a large scale height close to the star. These structures appear to be optically thin, but they still absorb and reprocess a significant fraction of the stellar light, which requires the dust to be distributed in a structure with high scale height, and not compressed toward the midplane of the disk. Such a large height of the structures is puzzling, since disks that are in hydrostatic equilibrium should be quite flat close to the star \citep{kenyon87,chiang97}. In this paper we study the possibility that the large height of the halo-like structure may be reached in a dynamical way by scattering planetesimals. This idea is inspired by the fact that planetesimals in the Solar System (in the Asteroid Belt and the Kuiper Belt) can have substantial inclinations, caused by interactions with planets. If this mechanism is found to be capable of explaining the observed optically thin structures, their presence could be used to infer the existence of planets in these systems. This work will focus on the question of whether a inwardly migrating planet could be responsible for creating such an optically thin halo-like structure.

Section \ref{sec:halos} discusses three YSOs that show this new structure. Section \ref{sec:directscattering} discusses high inclinations in our own Solar System, and quantitatively formulates a mechanism that could cause highly-inclined orbits in a system with one planet. In section \ref{sec:simulations} numerical simulations of the effect of a migrating planet on a planetesimal population are presented. The grinding down of the resulting planetesimal population to form small dust grains is covered in section \ref{sec:dust}. Finally, a discussion and conclusions are formulated in sections \ref{sec:discussion} and \ref{sec:conclusions}.

\section{Optically Thin Dust}\label{sec:halos}
Recently, a number of transitional disks have indications of possessing optically thin dust with large scale height: \object{LkCa 15}, \object{HD 142527} and \object{HD 163296}. These systems are discussed below.

	\subsection{LkCa 15}
\object{LkCa 15} is a low-mass pre-main-sequence star located at about 140 pc in the Taurus star-forming region \citep{espaillat07}. \citet{espaillat07} compared different models to SEDs obtained with the Infra-red Array Camera (IRAC) aboard Spitzer to study \object{LkCa 15}. An optically thick outer disk is found to have an inner radius of 46 AU, containing about $0.1 M_{\odot}$. The inner region of the disk is a bit more complex and \citeauthor{espaillat07} provide two options which could explain the shape of the SED in the near-IR: an optically thick disk from 0.12 to 0.15 AU combined with $4\times10^{-11}M_{\odot}$ of optically thin dust between 0.15 and 4 AU, or, $5\times10^{-11}M_{\odot}$ of optically thin dust between 0.12 and 4 AU. The first option is said to give the better fit. Both scenarios include an optically thin part, but it is stressed no dust can exist further out than $\sim5$ AU since the contribution at $20\rm \, \mu m$ becomes too strong \citep{espaillat07}. 

In 2009, \citeauthor{mulders10} investigate the inner region of \object{LkCa 15} in more detail and find an optically thick inner disk will influence the outer disk as well. Because the optically thick inner disk will obscure the outer disk, the outer disk needs to be blown up in the vertical direction in order for it to fit the observed SED. Alternatively, the inner disk could be optically thin, but this is only possible if the material is distributed over a large solid angle, virtually making it a halo. \citet{mulders10} used the 2D radiative transfer code MCMax \citep{min09} to model the observed SED. \citeauthor{mulders10} conclude it is hard to explain the large scale height of the inner dust material. One offered explanation is the violent scattering and disruption of planetesimals, which would end up in them colliding and filling the inner region with dust.

\citet{espaillat10} use new observations of the near-IR excess and find it to have the spectral shape of a blackbody, leading them to prefer the option of an optically thick inner disk and a blown-up outer disk. The inner disk is found to lie within 0.19 AU and the upper limit for the mass it contains is found to be $2\times10^{-4}M_{\odot}$. 

Whether the inner disk of \object{LkCa 15} contains an optically thin dust component remains a matter of debate, but if optically thin dust is responsible for the near-IR flux, the scale height of the material has to be very large. 

	\subsection{HD 142527}
Another interesting system is that of \object{HD 142527}. The star was categorized as an F6III star by \citet{houk78} and later classified as a Herbig star \citep{waelkens96}. \object{HD 142527} is about 20 times as luminous as the sun, has a mass of $2.2\pm0.3 M_{\odot}$, and shows a huge IR-flux ($F_{\rm IR}=0.92F_{*}$) which indicates there must be a considerable amount of circumstellar material \citep{verhoeff11}. 

\citeauthor{verhoeff11} used MCMax to model the circumstellar disk and find a model which accurately describes Spitzer and ISO IR spectra. A complicated structure is found for the system; an inner and an outer disk as well as an optically thin halo-like structure. The inner disk stretches from 0.3 to 30 AU and has a puffed up inner rim which shadows the rest of the disk. The small dust grains in the inner disk have a total mass of $2.5\times10^{-9}M_{\odot}$. It is argued most of the other material is already locked up in planetesimals or even planets. The degree of sedimentation also indicates an advanced state of evolution \citep{verhoeff11}.

Then there is the optically thin halo, which stretches from 0.3 to 30 AU and has an estimated dust mass of $1.3\times10^{-10}M_{\odot}$. The origin of this halo is unclear and again the dynamical scattering of planetesimals is offered as a speculative explanation. 

There is a gap in the disk stretching from 30 to 130 AU. \citeauthor{verhoeff11} explain why photo-evaporation is unlikely to be the cause of this gap and speculates about the presence of Jupiter or Earth-like planets as the origin of the gap. Lastly there is the outer disk, located between 120 and 200 AU, and with a rim height of 60 AU. Such a large rim height can be reached in hydrostatic equilibrium, but only if the disk is illuminated by the nearly unattenuated stellar radiation. A physical explanation for the high rim therefore requires the warm dust to remain optically thin and, therefore, to be spread out to large scale heights. The total mass in small dust grains is estimated to be $1.0\times10^{-3}M_{\odot}$.

	\subsection{HD 163296}
\object{HD 163296} is an isolated Herbig Ae star located at about 122 pc, which shows a NIR excess indicating a circumstellar disk \cite{hillenbrand92}. In 2010, \citeauthor{benisty10} present long-baseline spectro-interferometric observations using the AMBER instrument on VLTI. They find that the inner rim of a disk alone cannot reproduce their observations in the NIR and need an additional emission component between 0.10 and 0.45 AU, with a temperature of 1600 K and an optical depth of $\sim0.2$. 

\citeauthor{benisty10} discusses whether hot gas could be held responsible for the emission between 0.10 and 0.45 AU but conclude that this is probably not the case. They argue that more detailed, non-LTE models that include the transition from optically thin to optically thick layers in a dust-free environment are needed to completely rule out the possibility of hot gas being the main contributor to the NIR flux.

The existence of refractory dust grains is explored as another possibility for the emission. An optically thin layer of dust grains is assumed to lie between an inner and outer radius, which are both inside the inner rim of the optically thick disk. The vertical optical depth of the dust is assumed to be proportional to $1/r$, with $r$ the distance to the star. \citeauthor{benisty10} find the observed SED is reproduced best with the dust between 0.1 and 0.45 AU. The total mass in small dust between is calculated and found to be $8.7\times10^{-8}M_{\oplus}$ for Graphite grains of $0.05-2\rm \, \mu m$, and $9.5\times10^{-7}M_{\oplus}$ for Iron grains of $0.02-2\rm \, \mu m$ in size.

\section{Direct Scattering to Reach Highly-Inclined Orbits}\label{sec:directscattering}

  \subsection{Highly-Inclined Orbits in our Solar System}\label{sec:solarsystem}
How do we build a population of planetesimals on highly-inclined orbits from a nearly flat transitional disk? Hints for a mechanism can be found in the Kuiper Belt in our own Solar System. The Kuiper Belt can be split up into three dynamical classes \citep{jewitt00}: The classical Kuiper Belt, the resonant Kuiper Belt, and the scattered Kuiper Belt. Classical Kuiper Belt Objects (KBOs) have low-eccentricity, low-inclination orbits beyond 40 AU. The resonant KBOs are trapped in a mean motion resonance (MMR) with Neptune, and the scattered ones have probably suffered a close encounter with Neptune in the past and show a wide spread in inclination.  Scattered KBOs tend to have eccentric orbits with large semi-major axes. Studies on the distribution of inclinations of bodies in these three populations \citep{brown01,gulbis10} show they truly are distinct classes and it is argued that the inclinations in the three classes have different origins. 

Observational studies show resonant KBOs to have inclinations up to $30^{\circ}$ \citep{brown01,gulbis10}. They are commonly believed to have been trapped in an MMR with Neptune when the planet was migrating outward, having their eccentricity and inclination raised during the process. This mechanism explains Pluto's peculiar orbit \citep{malhotra93}, and inclinations of up to $\sim15^{\circ}$ in general for bodies outside Neptune's orbit \citep{malhotra95}. However, only a small fraction of the resonant KBOs in the simulations have their inclinations raised in excess of 10\% \citep{malhotra00}. Other mechanisms have been proposed to account for the observed inclinations, including sweeping secular resonances \citep{li08}, and stirring by large planetesimals \citep{morbidelli97}, but no definitive answer has been given.

Scattered KBOs show a wider spread in semi-major axes and their inclinations are usually greater than $10^{\circ}$ with a peak around $20^{\circ}$ \citep{gulbis10}. It is this broad distribution in orbital parameters that indicates that their orbits are probably the result of a close encounter with Neptune \citep{brown01}. \citet{hahn05} conclude from simulations that about 90\% of the KBOs in the scattered disk might not have suffered a close encounter, but rather obtained their orbits from Neptune's resonances during the migration epoch. Their model is unable to reproduce the observed abundance of KBOs with $i>15^{\circ}$ however. 

From studying the Kuiper Belt, it appears there are 2 clear mechanisms that are able to increase an object's inclination drastically; resonance trapping and close encounters. The degree of inclination-raising by resonance trapping is determined, and therefore limited by, the migration distance of the planet. \citet{malhotra95} shows a migration Neptune over a distance of 7 AU to cause an increase in eccentricity of 0.25 and in inclination of about $10^{\circ}$. The inclination that can be reached by direct scattering is limited by the relative velocity between the planet and the object and the escape velocity at the planet's surface, as we shall see in section \ref{sec:scattering}. It appears direct scattering is a more promising way to get bodies onto highly-inclined orbits, and it is this path that we will pursue in the rest of this work.

  \subsection{Direct Planetesimal Scattering}\label{sec:scattering}
Suppose we have a planet of mass $M_{\rm pl}$ and radius $R_{\rm pl}$ orbiting a star of mass $M_{*}$, on a circular orbit with Keplerian velocity $v_{\rm K}=\sqrt{GM_{*}/a_{\rm pl}}$. We assume a planetesimal of mass $m\ll M_{\rm pl}$ on an orbit with negligible inclination is approaching the planet with an impact parameter $b$ and a relative velocity $v_{\rm rel}$.

If $b$ is small enough, the close encounter can be treated as a 2 body interaction, without having to consider the central star. During this interaction, the magnitude of the \emph{relative} velocity is conserved but the direction is changed by a scattering angle of \citep[e.g.][]{weidenschilling75}
\begin{equation}\label{eq:theta1}
\theta = 2\arctan{\left[\frac{GM_{\rm pl}}{v^2_{\rm rel} b}\right]}.
\end{equation}
We normalize distances to the radius of the planet and velocities to the Keplerian velocity
\begin{equation}
 \hat{v}_i = \frac{v_i}{v_{\rm K}}, \hat{b} = \frac{b}{R_{\rm pl}}.
\end{equation}
This allows us to rewrite eq. \ref{eq:theta1} as follows
\begin{equation}\label{eq:theta2}
\theta = 2\arctan{\left[\frac{1}{2}\left(\frac{\hat{v}_{\rm esc}}{\hat{v}_{\rm rel}}\right)^2 \hat{b}^{-1}\right]},
\end{equation}
where $v_{\rm esc} = (2GM_{\rm pl}/R_{\rm pl})^{1/2}$ is the escape velocity at the surface of the planet. 

This equation shows that a scattering angle of $90^{\circ}$ is reached for 
\begin{equation}\label{eq:90deg}
 \left(\frac{v_{\rm rel}}{v_{\rm esc}}\right)^{2} = \frac{1}{2\hat{b}}.
\end{equation}
Thus, for velocities $v_{\rm rel}>v_{\rm esc}/\sqrt{2}$, an impact parameter smaller than the physical size of the planet is required. Such a planetesimal will crash into the planet. Due to gravitational focussing any planetesimal with an impact parameter smaller than
\begin{equation}\label{eq:bmin}
\hat{b}_{\rm min} = \sqrt{1 + \left(\frac{\hat{v}_{\rm esc}}{\hat{v}_{\rm rel}} \right)^{2}},
\end{equation}
will be lost in a physical collision with the planet \citep{safronov66}.

Thus, for a planet to be able to significantly scatter a planetesimal out of the orbital plane, the relative velocity of the encounter needs to be about a factor 2 lower than the escape velocity at the planet's surface.

Suppose the planet is able to scatter the planetesimal out of the orbital plane by an angle of $90^{\circ}$. In the frame of the planet, the planetesimal's total velocity is now pointing perpendicular to the orbital plane. In the stellar frame however, the planetesimal is also moving with the planet at the Keplerian velocity, and we can calculate the new inclination using
\begin{equation}
 i = \arctan\left[\frac{v_{\rm rel}}{v_{\rm K}}\right].
\end{equation}
This simple calculation shows that an inclination of $45^{\circ}$ can be reached for $v_{\rm rel}=v_{\rm K}$. 

Summing up, planetesimals can only be scattered into highly-inclined orbits for a particular combination of $v_{\rm rel}$, $v_{\rm esc}$ and $v_{\rm K}$. This condition can be written as
\begin{equation}\label{eq:velocities}
 v_{\rm K} \lesssim v_{\rm rel} \lesssim \frac{v_{\rm esc}}{2}.
\end{equation}

We can write the relative velocity of the collision as a function of orbital parameters of both objects. Recall that the planet moves on a circular orbit with semi-major axis $a_{\rm pl}$. We define the orbit of the planetesimal by its eccentricity $e$, semi-major axis $a$, and inclination $i$ relative to the planet. The relative velocity is then given by \citep{weidenschilling75}
\begin{equation}\label{eq:vrel}
 \hat{v}_{\rm rel}^{2} = 3 - 2\sqrt{\frac{a}{a_{\rm pl}}(1-e^{2})} \cos i - \frac{a_{\rm pl}}{a}.
\end{equation}
Recalling $i=0$ and taking $a=a_{\rm pl}$, we need an eccentricity of $e\geq(3/4)^{1/2}\approx0.87$ to get $v_{\rm rel}\geq v_{\rm K}$. Particles on orbits larger than that of the planet reach similar relative velocities for lower inclinations. However, if the particle is located inside the planetary orbit, for instance in the 1:2 MMR, it has $a/a_{\rm pl}\approx0.63$ and only reaches $v_{\rm rel}=v_{\rm K}$ for an eccentricity of 0.97. 

Since this study focusses on \emph{inward} planetary migration, most of the material will have $a<a_{\rm pl}$. It is obvious from eq. \ref{eq:vrel} that this means that the planetesimals need to be on (very) eccentric orbits. Section \ref{sec:rescap} will explore an explanation of these highly eccentric orbits. 

\begin{figure}
\begin{center}
\resizebox{\hsize}{!}{\includegraphics{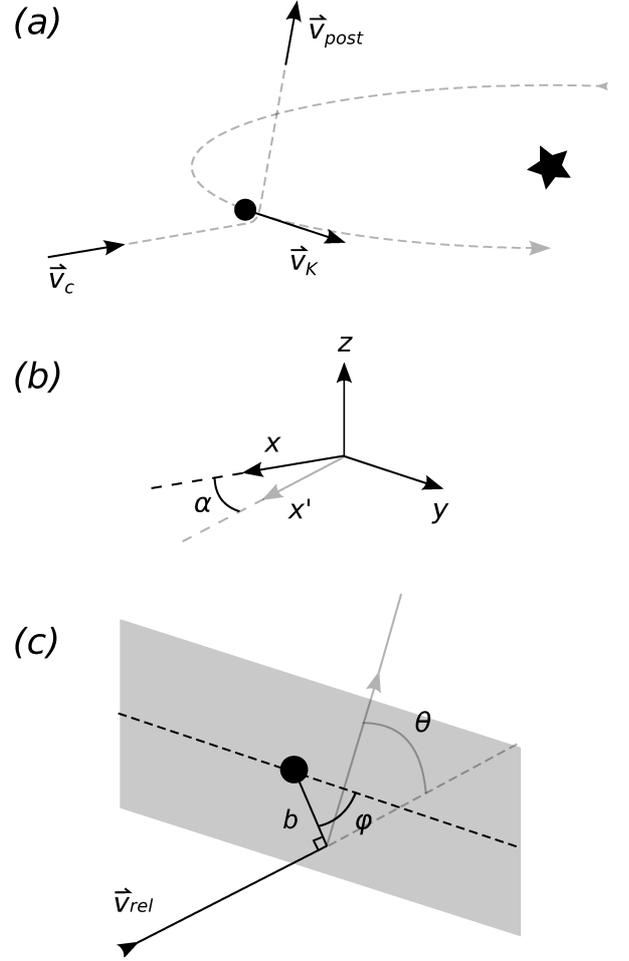}}
\caption{(a) Scattering of a planetesimal in the stellar frame. The planetesimal comes in along the \emph{x}-axis and the planet moves in the positive \emph{y}-direction. The \emph{z}-axis is perpendicular to the orbital plane. (b) Unit vectors \emph{x} and \emph{x'} (and \emph{y} and \emph{y'}) lie in the orbital plane, but are rotated with respect to each other by an angle $\alpha$. (c) The scattering in the frame of the planet. The planetesimal comes in along the \emph{x'}-axis, with impact parameter $b$. It is scattered in the (\emph{y'-z})-plane over an angle $\theta$.}
    \label{fig:coords}
  \end{center}
\end{figure}

The remainder of this section will calculate post-encounter inclinations in more detail and study the conditions for which planetesimals are scattered out of the system on hyperbolic orbits. For simplicity, we assume the orbits of the planetesimals to have $e=1$, causing the velocities of the planet and the planetesimal to be perpendicular at the point of close encounter. Figure \ref{fig:coords} shows the different quantities involved. Figure \ref{fig:coords}a depicts the stellar frame. In this case the comet velocity points in the negative \emph{x}-direction, the planet moves in the direction of the positive \emph{y}-axis, and the \emph{z}-axis is perpendicular to the orbital plane. 

Figure \ref{fig:coords}c shows the scattering in the frame of the planet. In this frame the planetesimal's velocity points in the negative $x'$-direction, and has a size $v_{\rm rel}$ determined by 
\begin{equation}\label{eq:vrel2}
\hat{v}_{\rm rel} =\sqrt{1 + \hat{v}_{\rm c}^2}.
\end{equation}
The shaded area depicts the plane in which the scattering over an angle of $\theta$ occurs, which is the (\emph{y'-z})-plane. Unit vectors \emph{x} and \emph{x'} (and \emph{y} and \emph{y'}) lie in the orbital plane, but are rotated with respect to each other by an angle of 
\begin{equation}
\alpha = \arctan(\hat{v}_{\rm c}).
\end{equation}
as depicted in figure \ref{fig:coords}b. 

As a planetesimal comes in (along the \emph{x'}-axis) with impact parameter $\hat{b}$, the scattering angle $\theta$ is determined by eq. \ref{eq:theta2}. We repeat that a comet approaching with $\hat{b}<\hat{b}_{\rm min}$ will have a physical collision with the planet.

For the post-encounter comet\footnote{The words \emph{comet} and \emph{planetesimal} are used throughout this paper to describe bodies of mass $m\ll M_{\rm pl}$. \emph{Comet} is mostly used to refer to a particle on an orbit with a large eccentricity.} inclination ($i$) and velocity ($v_{\rm post}$), it is not just the size of the impact parameter that matters, but also the direction of this offset with respect to the motion of the planet. To quantify this direction we use $\phi$. If the planetesimal is aiming at a point slightly behind the planet ($\phi=\pi$), it will be scattered in the planet's direction of motion, and therefore accelerated. If the planetesimal is aiming at a point directly above or below the planet ($\phi=\pm\pi/2$), it is scattered out of the orbital plane.

If we split the post-encounter velocity of the comet into a component in the \emph{z}-direction and one in the orbital plane, $\hat{v}_{\rm post,z}$ and $\hat{v}_{\rm post,xy}$ respectively, we can calculate the total velocity to be
\begin{equation}\label{eq:vnew}
 \hat{v}_{\rm post} = \sqrt{\hat{v}_{\rm post,xy}^2 + \hat{v}_{\rm post,z}^2},
\end{equation}
and the post-encounter inclination using
\begin{equation}\label{eq:i}
 i = \arctan{\left[\frac{\hat{v}_{\rm post,z}}{\hat{v}_{\rm post,xy}}\right]}. 
\end{equation}
Since the escape velocity from the central star is equal to $\sqrt{2}v_{\rm K}$, comets with $\hat{v}_{\rm post}>\sqrt{2}$ will leave the system on a hyperbolic orbit.

For a given combination of $\hat{v}_{\rm c}$, $\hat{v}_{\rm esc}$, $\hat{b}$ and $\phi$, the resulting inclination and total velocity are uniquely defined and can be calculated using the following prescription:
\begin{enumerate}
 \item calculate $\hat{v}_{\rm rel}$ and $\hat{b}_{\rm min}$ from eq. \ref{eq:vrel2} and \ref{eq:bmin},
 \item determine angles $\alpha$ and $\theta$,
 \item in the planet frame, rotate $\hat{v}_{\rm rel}$ using $\theta$ and $\phi$,
 \item switch to the stellar frame by adding the planet's velocity under the correct angle $\alpha$,
 \item calculate $\hat{v}_{\rm post}$ using eq. \ref{eq:vnew} and check if the particle is ejected,
 \item find the comet's post-encounter inclination from eq. \ref{eq:i}.
\end{enumerate}

\begin{figure*}
\begin{center}
\resizebox{\hsize}{!}{\includegraphics{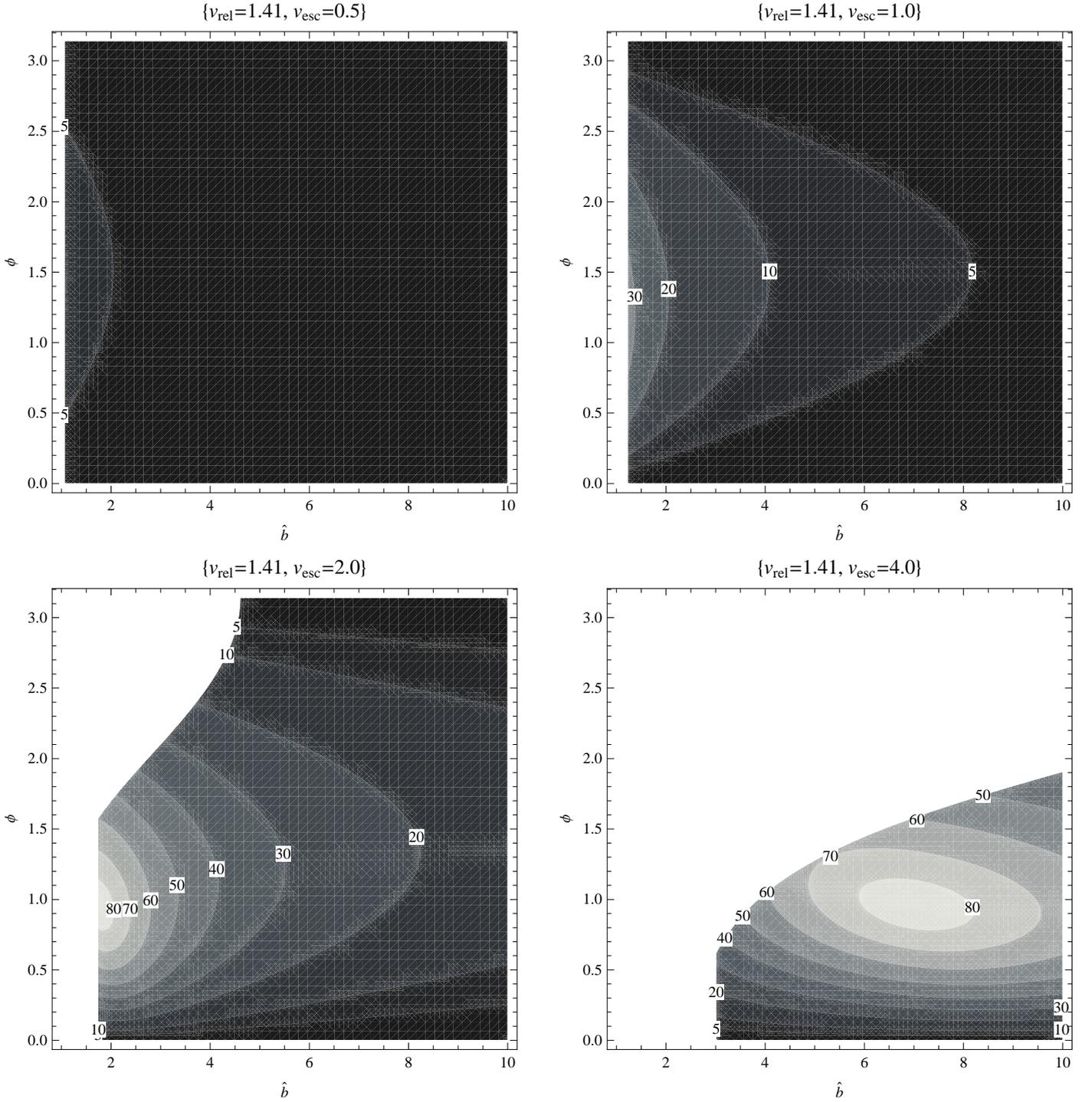}}
\caption{The inclination of a comet after an encounter at $\hat{v}_{\rm rel}=1.41$ with a planet with $\hat{v}_{\rm esc}$, as a function of impact parameter and $\phi$. The white regions mark the combinations of $\hat{b}$ and $\phi$ where $\hat{v}_{\rm post}>\sqrt{2}$. Regions where $b<b_{\rm min}$ are also coloured white.}
    \label{fig:inc}
  \end{center}
\end{figure*}

Figure \ref{fig:inc} shows the post-encounter inclination of a comet coming in with $\hat{v}_{\rm c}=1.41$ as a function of $\hat{b}$ and $\phi$ for four different values of $\hat{v}_{\rm esc}$. The regions where the comet is lost because of ejection from the system or a collision with the planet are coloured white. 

For a given system, ejection is more likely to happen for comets with relatively small impact parameters, and $\phi \sim \pi$. The latter is caused by these comets coming in behind the planet, and thus being accelerated in the stellar frame during the close encounter. The value of $\phi$ that is ideal for reaching high inclinations appears to lie between 0 and $\pi/2$, because the comet not only gains velocity in the \emph{z}-direction, but it is also slowed down in the \emph{xy}-plane.

The planet with the lowest mass can be seen not to be able to eject bodies, nor can it scatter them to orbits with inclinations $>5^{\circ}$. As the escape velocity at the planet's surface is increased, three things happen; 1) $\hat{b}_{\rm min}$ increases, 2) the region where particles are ejected grows larger (both in $\phi$ and in $\hat{b}$), and 3) the region where high inclinations are reached shifts towards higher impact parameters. Figure \ref{fig:inc} confirms equation \ref{eq:velocities}, as an escape velocity higher than the relative velocity is needed to scatter bodies to highly-inclined orbits. 

  \subsection{Eccentricity Pumping and Migration}\label{sec:epumpingandmigration}
The relative velocity used in fig. \ref{fig:inc} can only be reached for very eccentric planetesimal orbits. A possible explanation for high eccentricities is resonance trapping, see for instance \citet{malhotra93,malhotra95,malhotra00}. In our own Solar System, the trapping and consequent dragging along of Plutinos in Neptune's 3:2 resonance, has raised their inclinations from nearly 0 to up to $30^{\circ}$ (see section \ref{sec:solarsystem}).

A planetesimal orbiting a star is in an MMR with a planet if their mean motions are locked, and the resonant argument is librating (see \citet{murray99} for a detailed treatment of the resonant argument). The locations of the mean motion resonances of a planet orbiting a star at $a_{\rm pl}$ are
\begin{equation}\label{eq:mmr}
 a_{(p+q):p}=a_{\rm pl}\left(\frac{p+q}{p}\right)^{2/3},
\end{equation}
with $p$ and $q$ integers, and the ratio of an MMR depicting the ratio of the orbital periods, so that a 1:2 resonance lies inside the planet's orbit while a 2:1 resonance lies outside. Resonances with $|q|=1$ are called first-order resonances and are often described with $j$ instead of $p$.

When the planet is migrating, the locations of the MMRs will change and bodies trapped within a resonance can be dragged along. This will not only change the dragged-along body's semi-major axis, but also various other orbital parameters. \citet{malhotra93} showed Pluto's peculiar orbit could be explained by it being stuck in Neptune's 3:2 resonance, as Neptune migrated out by $\sim7$ AU. 

\citet{wyatt03} provides us with an expression for the eccentricity of a planetesimal orbit, that has been dragged in a $(p+q):p$ resonance by a planet 
\begin{equation}\label{eq:epump}
 e^2 = e_{0}^{2} + \left(\frac{q}{p+q}\right)\ln{\frac{a}{a_{0}}},
\end{equation}
where $a_{0}$ and $a$ are the initial and current semi-major axes of planet. Supposing a planetesimal gets captured in the 2:1 resonance with $e_{0}=0$, it would have to be dragged to $a/a_{0}\backsimeq5$ to reach an eccentricity of 0.9.

Resonance capture is not guaranteed however. As the resonance of a planet moves in (or out), because of the planet migrating, and it encounters a planetesimal, the probability of capture depends on both eccentricities, the migration rate, and planet mass \citep{mustill10}. \citeauthor{mustill10} conclude resonance capture is guaranteed for low eccentricity and low migration rates, impossible at fast migration rates and low eccentricities, and possible for fast migration rates and high eccentricities. The transitions between these regimes depend on the planet mass. Furthermore, the critical migration rate for capture increases with $j$, so if the particle survives the passing of the 2:1 resonance, it may still be captured in another resonance closer to the planet with higher $j$. It is also pointed out that planetesimals that are not caught, experience a small jump in $e$ up or down depending on the migration rate. \citet{mustill10} calculate capture probabilities as a function of migration rate for different MMRs, migration rates, and planetary masses. For instance, a Neptune-mass planet is found to capture all bodies in its 3:2 resonance for migration rates $\lesssim 4\times10^{-6}\rm \, AU\,yr^{-1}$.

Resonance trapping followed by eccentricity-pumping appears to be needed to ensure high relative velocities. Another argument for an inwardly migrating planet is the fact it can collect and carry material inwards. A stationary planet can only scatter planetesimals in its direct vicinity. Since the optically thin structures are located close to their central stars and have considerable mass in small dust grains (see section \ref{sec:halos}) this build-up of material by an inwardly migrating planet is crucial.

Assume a planetesimal disk with a mass distribution equal to the mass in solids in a minimum mass Solar Nebula (MMSN). The surface density of solids in a MMSN disk is given by \citep{hayashi81}
\begin{equation}\label{eq:mmsn}
 \Sigma_{\rm s}=\Sigma_{0}\left(\frac{r}{AU}\right)^{-3/2}\rm \,g\,cm^{-2}.
\end{equation}
$\Sigma_{0}=7.1$ for $r<2.7$ AU because ices cannot survive in the hot inner regions. Outside the snowline $\Sigma_{0}=30$. 

A stationary planet of Neptune-mass on a circular orbit at 5 AU can scatter planetesimals within a region the size of its Hill sphere \citep[e.g.][]{armitage10}
\begin{equation}\label{eq:rhill}
 \Delta r = R_{\rm H} = a\left(\frac{M_{\rm pl}}{3M_{*}}\right)^{1/3}. 
\end{equation}
The mass of planetesimals within a ring of this size is given by
\begin{equation}\label{eq:deltam}
 \Delta m = 2\pi a \Sigma_{\rm p} \Delta r.
\end{equation}
Using $\Sigma_{\rm p}=30\rm \,g\,cm^{-2}$, we find $\Delta m \simeq 4.6M_{\oplus}$.

Assume now that the planet migrated from 30 to 5 AU at such a rate, it collected all material in between. We can integrate eq. \ref{eq:mmsn} to find $\Delta m\simeq46M_{\oplus}$. This is an increase by a factor of 10 with respect to the mass scattered by a stationary planet at 5 AU. The difference becomes even bigger when looking at planets closer to the star, since $\Delta m$ in eq. \ref{eq:deltam} is proportional to $a^{2}$.

\section{Simulations}\label{sec:simulations}

	\subsection{Introduction}
\begin{figure}
\begin{center}
\resizebox{\hsize}{!}{\includegraphics{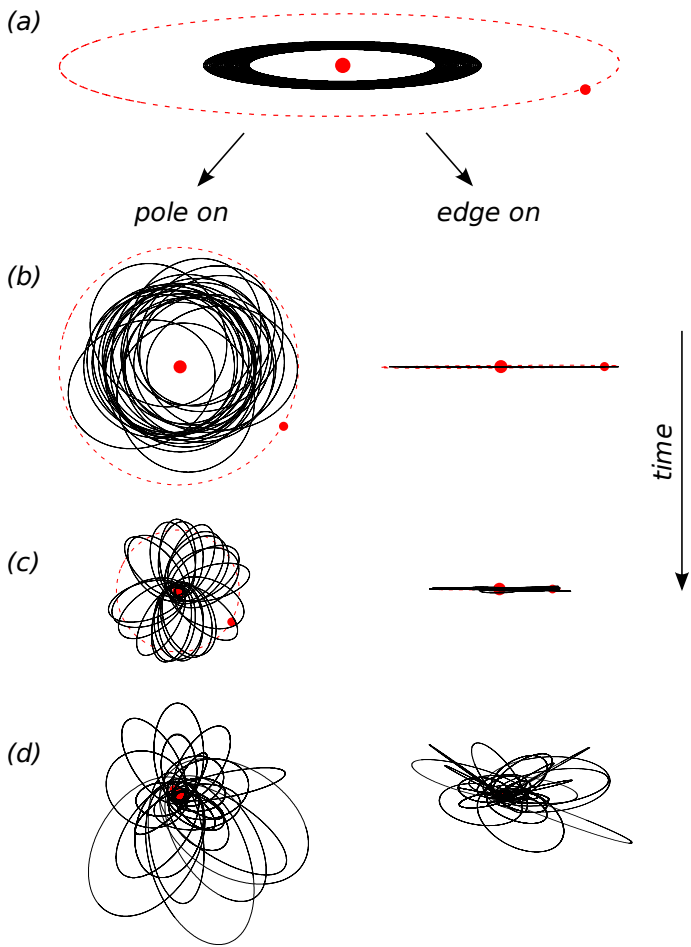}}
\caption{Illustration of the mechanism to reach a planetesimal population with high orbits, as explored in this paper. (a) both the planet orbit (dotted) and planetesimal orbits (solid) start out in the orbital plane. (b) As the planet has migrated in a some distance, the planetesimals are stuck in an MMR. As they are dragged along their orbits become increasingly eccentric. (c) At some point the planetesimal orbits become so eccentric, the bodies become planet-crossing. As the eccentricity rises even further, the relative velocity between the planet and the planetesimals grows. (d) The high relative velocity compared to the local Keplerian velocity enables the planet to violently scatter the planetesimals throughout the system, to orbits with high inclinations.}
    \label{fig:sketch}
  \end{center}
\end{figure}

The scenario for creating the highly-inclined orbits is now as follows. A planet with $v_{\rm esc}\gtrsim v_{\rm K}$ and a planetesimal population start out on (nearly) circular orbits and in the orbital plane (figure \ref{fig:sketch}a). Due to for example the presence of gas the planet starts migrating inwards, and collecting material in mean motion resonances. Planetesimals that are stuck in an MMR, have their eccentricities raised significantly. Figure \ref{fig:sketch}b shows the system after some migration has taken place. All planetesimal orbits in this figure have the same semi-major axis (they are presumed stuck in the same MMR) but differ in eccentricity. As the planet migrates even further, the increasing eccentricities cause the orbits of the planetesimals to be planet-crossing (figure \ref{fig:sketch}c). At this point, the inclinations are only a couple of degrees. The relative velocity between the planet and the planetesimals grows together with the eccentricity of the planetesimal orbit and at some point this will lead to violent scattering events. As the planet has nearly reached the star it will leave behind a scattered population of planetesimals, which show (depending on the planet's escape velocity) a large spread in inclinations, figure \ref{fig:sketch}d.  This population of scattered planetesimals, over time, will collide and grind down to form the small dust that causes the optical depth. In this section numerical simulations are presented which show the resonance capture, eccentricity-pumping, and scattering of planetesimals happening simultaneously.

\subsection{Setup}\label{sec:setup}
For the integrations of the migrating planet the software package Swifter is used\footnote{http://www.boulder.swri.edu/swifter/}. Swifter is written by D.E. Kaufmann and based upon the Swift package developed by Levison \& Duncan. Within Swifter, the user is given the choice between 7 integrating schemes, including Wisdom-Holman Mapping \citep{wisdom91}, the Regularized Mixed Variable Symplectic (RMVS) integrator \citep{levison94}, and the Symplectic Massive Body Algorithm (SyMBA) \citep{duncan98,levison00}. In addition, Swifter also includes a Bulirsch-Stoer integrator \citep{teukolsky92}. In this work SyMBA was used most of the time, since it can handle close encounters correctly and is significantly faster than the Bulirsch-Stoer method. The results of the SyMBA integrator are checked by comparing a few cases to a Bulirsch-Stoer integration with a very small timestep.

During the integrations, a planet is pushed to migrate inwards from 30 AU, while a population of 100 massless planetesimals orbits the star at circular orbits between 10 and 15 AU. The planetesimals all start out with $e=i=0$, whereas the planet is given a small inclination of $0.72^{\circ}$, to ensure $\phi$ can take on all values between 0 and $2\pi$, and an eccentricity of $0.012$. The planet mass is varied between 0.1 and 20 Neptune masses. 

For simplicity, the migration is not done in a self-consistent fashion. We want to focus on the effect of the migrating planet on the planetesimal population, to see if the resonance capture, eccentricity pumping, and scattering can take place. A self-consistent migration would vastly increase the complexity of the problem. Instead, the planet is pushed within SyMBA to have it migrate at a constant rate $\xi$, which is varied between $7.3\times10^{-6}\rm \, AU\,yr^{-1}$ and $7.3\times10^{-4}\rm \, AU\,yr^{-1}$. These migration rates are based upon estimates for migration rates and time scales for type I migration \citep{tanaka02}, type II migration \citep{armitage10}, and migration through a planetesimal disk \citep{armitage10}. Since the migration in this work is artificial, the system does not contain any gas or massive planetesimals and, therefore, does not distinguish between the various types of migration. Section \ref{sec:discussion} will discuss what effect the presence of gas or other drivers for migration might have on the results. 

For SyMBA to be able to handle close encounters between massless bodies and planets, a Hill-sphere radius, $R_{\rm H}$, needs to be specified for every planet. This radius determines when SyMBA will shorten the integration timestep. Since the Hill-sphere radius is proportional to a planet's semi-major axis (eq. \ref{eq:rhill}), it will decrease in time for a planet that migrates inwards. For simplicity, we will treat $R_{\rm H}$ as a constant. This results in the size of the Hill sphere being overestimated as the planet migrates in, slowing down the integrations somewhat.

The system is evolved long enough for the planet to pass through the planetesimal population, and end up close to the star, $<1$ AU, where it has a negligible effect on the planetesimals. The timestep for the integrations ranges between 0.5 days and $10^{-1}$ yr, depending on the smallest perihelion distance in the simulation, making sure all orbits are resolved. Within SyMBA, the timestep is decreased when a planetesimal enters the Hill sphere of the planet. During the simulations, planetesimals are removed if they
\begin{itemize}
\item come within a distance $r_{\rm min}$ from the origin. We use $r_{\rm min}=0.05$ AU.
\item have $a>r_{\rm max}$. In our simulations $r_{\rm max} = 1000$ AU is used.
\item come within $R_{\rm pl}$. In this work $R_{\rm pl}=R_{\rm Nep}$ is used.
\end{itemize}
The choice for $r_{\rm min}$ comes from the fact that very eccentric orbits with perihelion distances smaller than 0.05 AU are hard to resolve without using a very small timestep, and will lead to non-physical behaviour. Furthermore, bodies that come this close to the central star are likely to be evaporated.

	\subsection{Resonance Capture}\label{sec:rescap}
\begin{figure}
\resizebox{\hsize}{!}{\includegraphics{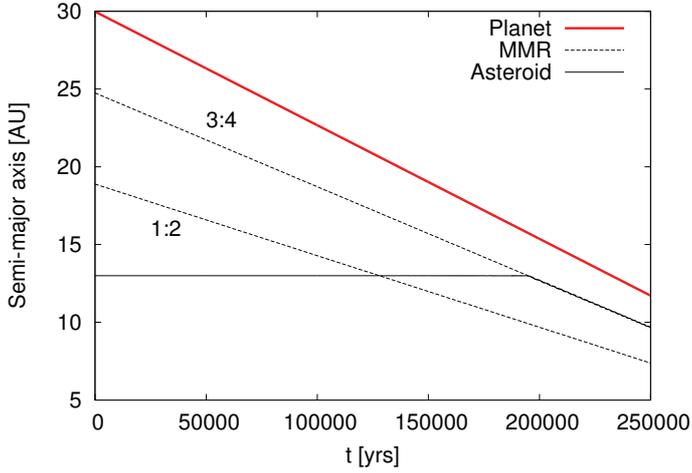}}
\caption{A Neptune-mass planet is migrating in with $\xi=7.3\times10^{-5}\rm \, AU\,yr^{-1}$. The asteroid survives the passing of the 1:2 resonance but is captured in the 3:4 resonance and dragged inward.}\label{fig:54_a_br}
\end{figure}

An example of resonance capture in the simulations is illustrated in fig. \ref{fig:54_a_br}. The figure shows a Neptune-mass planet being pushed artificially to migrate in from 30 AU at a rate of $7.3\times10^{-5}\rm \, AU\,yr^{-1}$. A massless asteroid is orbiting the solar-mass star at 13 AU on an initially circular orbit. The asteroid survives the passing of the 1:2 ($j=1$) MMR but is eventually captured by the 3:4 ($j=3$) MMR. For lower migration rates (or higher planet masses) the asteroid gets trapped in the 1:2 resonance already, while for higher migration rates (or lower planet masses) the asteroid only gets captured very close to the planet.

\begin{figure}
\resizebox{\hsize}{!}{\includegraphics{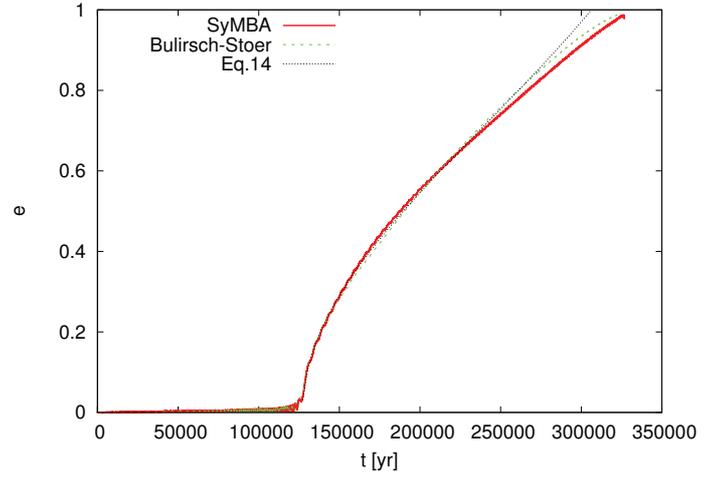}}
\caption{Evolution of the eccentricity of a massless body stuck in the 1:2 internal mean motion resonance of a $10M_{\rm Nep}$ planet migrating inward at a rate of $7.3\times10^{-5}\rm \,AU\,yr^{-1}$. The timestep for both the SyMBA and Bulirsch-Stoer integration was set at 0.5 days. The dotted line shows the expected increase in eccentricity as predicted by equation \ref{eq:epump}. The test particle is captured at $t\approx0.13$ Myr and eventually lost to the central star after about 0.32 Myr.}\label{fig:epumping}
\end{figure}

Figure \ref{fig:epumping} shows the evolution of a massless body as it is dragged along in a migrating planet's 1:2 resonance. The integration has been performed using two different integrators, to check the consistency of SyMBA for highly eccentric orbits. The test particle is captured at $t\approx0.13$ Myr and eventually lost to the central star after about 0.32 Myr. During this time, the $10M_{\rm Nep}$ mass planet has travelled $\sim14$ AU. The results of the SyMBA and Bulirsch-Stoer integrators are very similar, and in both cases the massless body is eventually lost to the central star when it reaches $e\approx0.99$. The expected increase in eccentricity, as predicted by eq. \ref{eq:epump} is also plotted and seen to agree well with the simulations. The discrepancy between eq. \ref{eq:epump} and the simulations towards higher $e$ comes from the fact eq. \ref{eq:epump} is a first-order approximation.

    \subsection{Inclinations}\label{sec:inclinations}
\begin{figure*}
\begin{center}
\resizebox{0.8\hsize}{!}{\includegraphics{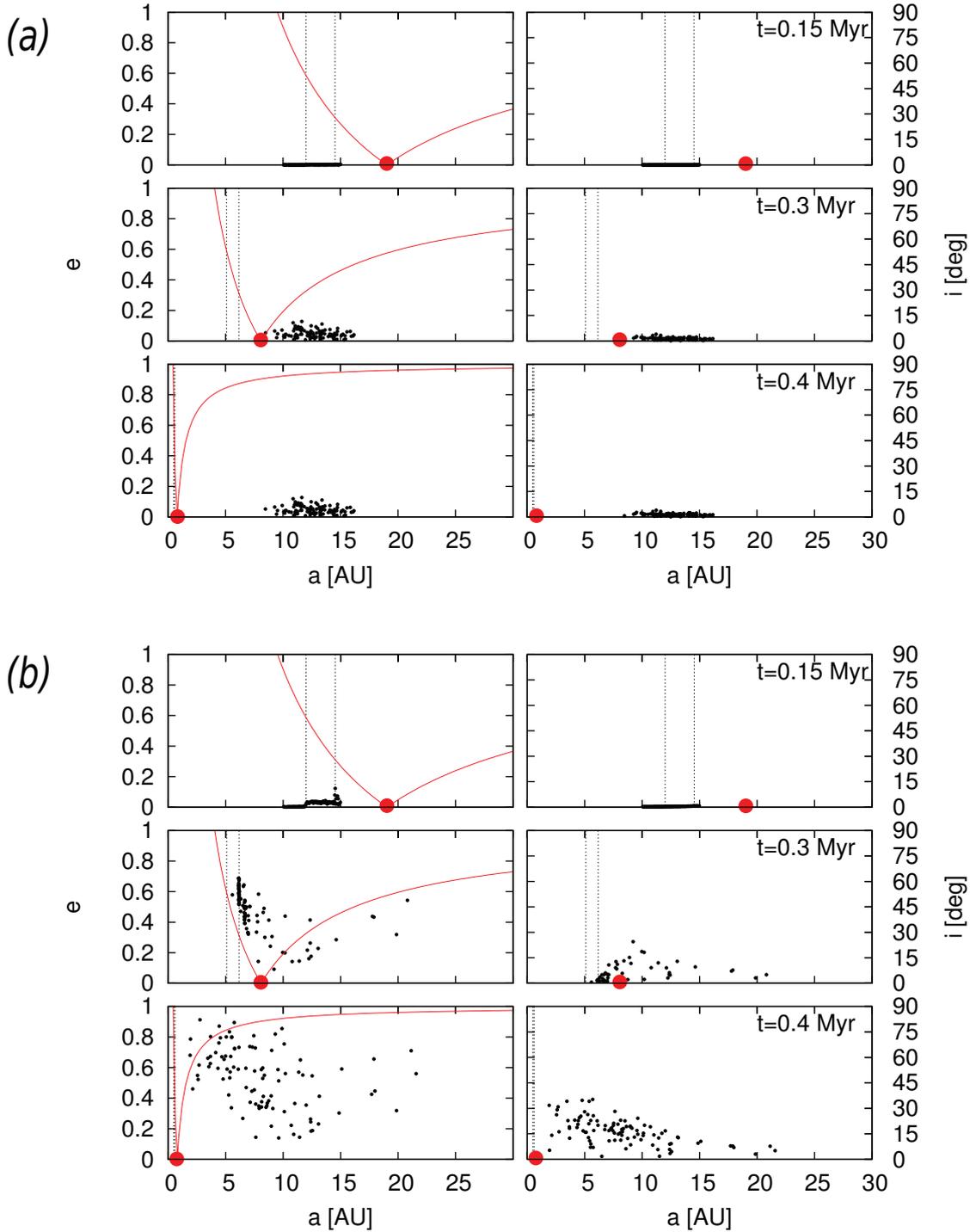}}
\caption{Evolution of the eccentricities (left) and inclinations (right) in a planetesimal swarm as a planet (large circle) migrates inwards from 30 AU at a rate of $7.3\times10^{-5}\rm \, AU\,yr^{-1}$. The planet mass used are $0.1M_{\rm Nep}$ for (a), and $2M_{\rm Nep}$ for (b). Planetesimals (small circles) above the solid lines are on a planet crossing orbit, and the dotted lines show the location of the 1:2 and 2:3 MMR.}
    \label{fig:all_BW}
  \end{center}
\end{figure*}

Figure \ref{fig:all_BW}a shows the evolution of the planetesimal swarm as a function of time. The left panels show eccentricity, the right panels inclination, and time passes from top to bottom. The large circle shows the location of the planet and the small circles represent massless planetesimals. Planetesimals above the solid lines are on a planet-crossing orbit. The dotted lines show the 1:2 and 2:3 MMR, where the 1:2 MMR is furthest from the planet. The planet is $0.1M_{\rm Nep}$ and migrating at a rate of $7.3\times10^{-5}\rm \, AU\,yr^{-1}$. The planetesimals start out between 10 and 15 AU with $e=i=0$. After 0.15 Myr (top panels), both resonances have reached the planetesimal population, and both appear unable to capture planetesimals. The passing of the resonances through the swarm of planetesimals is seen to have a very small effect on their orbital parameters. After 0.3 Myr the planet itself has migrated through the population of massless bodies, and has scattered most of them via direct encounters. Since the planetesimal orbits were not excited before the encounter, the relative velocity remained small, and the scattering events proved unable to get the planetesimal inclinations higher than a couple of degrees.The resulting population, shown after 0.4 Myr in the bottom panels, is very similar to the one after 0.3 Myr, i.e., the planet has little effect on the bodies once it has migrated inside the population. During this simulation, 2\% of the planetesimals is accreted by the migrating planet.
 
If we increase the planet mass, this will lead to a high capture probability. Figure \ref{fig:all_BW}b shows a similar calculation, but for a planet mass of $2M_{\rm Nep}$. After 0.15 Myr the effect of the enhanced planet mass are already visible as the passing of the 1:2 resonance raises the eccentricities of the planetesimals slightly, while the 2:3 resonance appears to be able to capture bodies, and subsequently drag them along. This becomes even more visible after 0.3 Myr, where a significant fraction of the planetesimals are stuck in the 2:3 MMR. As these bodies are dragged along, their eccentricities are raised as described by eq. \ref{eq:epump}. As the eccentricity of a planetesimal goes up, it will eventually reach a planet-crossing orbit, allowing it to have a close encounter with the planet. This has already happened for the bodies no longer in the 2:3 resonance. Since the close encounters happen when the planetesimals are on eccentric orbits, the relative velocities involved are higher, and the inclinations resulting from the scattering events are much greater than for figure \ref{fig:all_BW}a. During this simulation, 1\% of the bodies were accreted by the central star, and 1\% by the migrating planet.

If we increase the planet mass even further, the capture probability reaches almost unity, and the vast majority of planetesimals get stuck in a resonance far away from the planet. As they are dragged along and their eccentricities go up, they are protected from a close encounter with the planet for some time, since they are so far away. It turns out, almost all these bodies will reach $e\gg0.9$, and are accreted by the central star before they have time to be scattered by the migrating planet. An example of this scenario is shown in figure \ref{fig:epumping}. This means, that in the systems where you would expect to find the highest relative velocities (and therefore the highest inclinations) the majority of the planetesimals are actually lost to the central object. A way to get around this, might be to put an additional, stationary planet close to the central star, which will scatter the highly excited planetesimals before they come too close to the central object. This planet could not be too massive, as this would lead to it ejecting the planetesimals on hyperbolic orbits. In this study we have restricted ourselves to single-planet systems, but the study of stellar systems with multiple planets, one of which is migrating, might be very interesting indeed. Another way to protect the planetesimals from accretion by the central star would be to introduce a gas drag, which would dampen the eccentricities. The effect of gas of the results is further discussed in section \ref{sec:discussion}.

\begin{table*}
\caption{Orbital parameters of the resulting planetesimal population for a number of simulations.}
\label{table:simulations}
\centering
\begin{tabular}{c c c c c c c c c}
\hline\hline
Sim & $\xi$[$\rm AU\,yr^{-1}$] & $M_{\rm pl}/M_{\rm Nep}$ & $N_{\rm ej}[\%]$ & $N_{\rm acc}[\%]$ & $\langle a \rangle$[AU] & $\langle e \rangle$ & $\langle i \rangle[^{\circ}]$ & $i_{\rm max}[^{\circ}]$\\ \hline
S0.1&$7.3\times10^{-6}$&		0.1&	0&	0&	9.32&   0.255&  7.39&	16.16\\
S1&$7.3\times10^{-6}$&		1&	0&	100&	-&	-&	-&	-\\
M0.1&$7.3\times10^{-5}$&		0.1&	0&	0&	12.72&  0.047&  1.39&	3.98\\
M1&$7.3\times10^{-5}$&		1&	0&	0&	10.51&	0.311&	8.13&	19.00\\
M2&$7.3\times10^{-5}$&		2&	0&	1&	8.61& 	0.538&  14.61&	35.25\\
M10&$7.3\times10^{-5}$&		10&	0&	100&	-&  -&  -&	-\\
M20&$7.3\times10^{-5}$&		20&	0&	100&	-&  -&  -&	-\\
F0.1&$7.3\times10^{-4}$&		0.1&	0&	0&	12.61&  0.016&  0.29&	1.84\\
F1&$7.3\times10^{-4}$&		1&	0&	0&	12.59&  0.082&  1.84&	5.92\\
F2&$7.3\times10^{-4}$&		2&	0&	0&	13.23&  0.113&  2.50&	9.10\\
F10&$7.3\times10^{-4}$&		10&	0&	0&	12.30&  0.442&  9.60&	27.86\\
F20&$7.3\times10^{-4}$&		20&	0&	0&	18.79&  0.408&  9.42&	47.23\\
\hline
\end{tabular}
\tablefoot{$N_{\rm ej}$ and $N_{\rm acc}$ give the percentage of planetesimals that were lost during the integration resulting from respectively ejection from the system, or accretion by the central star. Inclinations are given in degrees. Simulation names consist of one letter describing the migration (F=Fast, M=Medium, S=Slow) and a number representing the planet mass.} 
\end{table*}

From section \ref{sec:rescap} we know that the capture probability also depends on the migration rate of the planet. We have therefore conducted a small parameter study, where the migration rate and planet mass have been varied. In total we carried out 12 simulations with the planet-mass ranging from $0.1-20M_{\rm Nep}$ and $\xi$ between $7.3\times10^{-4}-7.3\times10^{-6}\rm \, AU\,yr^{-1}$. Table \ref{table:simulations} shows the characteristics of planetesimal populations after migration has taken place, as well as the fraction of planetesimals which have been lost during the integration. Both higher planet mass and lower migration rate appear to ensure high average inclinations. This is the case because; 1) massive planets are more capable of scattering planetesimals into highly-inclined orbits (see section \ref{sec:scattering}), and 2) relatively massive planets and low migration rates increase the probability of resonance capture, allowing for more eccentricity-pumping and a higher relative velocity between the planetesimal and the planet during the close encounter. However, if the combination of migration rate and planet mass causes the planetesimals to get stuck in a resonance too far, they are likely to be lost to the central star before having a close encounter with the migrating planet.

\section{Dust Production}\label{sec:dust}
Section \ref{sec:simulations} has left us with populations of planetesimals and constraints on their orbital parameters. Section \ref{sec:colltimes} will calculate the time scales on which these planetesimals should collide and form dust, and section \ref{sec:dsd} will estimate the total mass and surface area of the dust size distribution resulting from the collisional cascade.

	\subsection{Collision Times}\label{sec:colltimes}
Suppose we have $N$ planetesimals of radius $R$, mass $m$ and a geometrical cross-section for collisions of $\sigma_{\rm coll} = 4\pi R^2$, flying around in some volume $V$. We can define the sweeping time as 
\begin{equation}
 t_{\rm s} = \frac{V}{v_{\rm coll}\sigma_{\rm coll}},
\end{equation}
where $v_{\rm coll}$ is the collision velocity between the planetesimals. Since the populations from table \ref{table:simulations} show pretty eccentric orbits, the collision velocity will vary a lot between collisions, but for the purpose of this calculation we estimate it as
\begin{equation}
 v_{\rm coll} = \eta v_{\rm K}\left(\langle a \rangle\right),
\end{equation}
with $\eta$ between 0.5 and 2. For two nearly circular orbits $\eta$ is determined by the vertical component of the individual velocities \citep{dominik03}. But since the orbits in our populations are eccentric and randomly oriented, $\eta$ will lie closer to $\sqrt{2}$.

If we assume the volume is wedge-shaped and lies between an inner and an outer radius, $r_{\rm in}$ and $r_{\rm out}$, and has a normalized height of $h=H/r$, we can write it as
\begin{equation}
 V = \frac{4\pi}{3}h(r_{\rm out}^3 - r_{\rm in}^3),
\end{equation}
where $h=\tan{i_{\rm max}}$. 

We have to think carefully about $r_{\rm in}$ and $r_{\rm out}$. One might insert the minimum and maximum values of the semi-major axis found in the resulting populations, but this is not correct since most comets are on pretty eccentric orbits. On the other hand, choosing $r_{\rm out}$ to be equal to $a_{\rm max}(1+e_{\rm max})$ will increase the volume drastically even though only one planetesimal is able to get to $r_{\rm out}$. It makes more sense to define something like
\begin{eqnarray}
r_{\rm out} &= (\langle a\rangle+\sigma_{a})(1+\langle e\rangle),\\
r_{\rm in} &= (\langle a\rangle-\sigma_{a})(1-\langle e\rangle),
\end{eqnarray}
where $\langle a\rangle$ and $\langle e\rangle$ can be taken from table \ref{table:simulations}. We approximate $\sigma_{a} \sim \sqrt{\langle a\rangle}$, which is found to work well for most populations in table \ref{table:simulations}.

The collision time for the comets will become
\begin{equation}
t_{\rm coll} = \frac{t_{s}}{N},
\end{equation}
with the number of bodies being constrained by the total mass of the population $M$
\begin{equation}
N=\frac{M}{m}=\left(\frac{R}{R_{\oplus}}\right)^{-3}\left(\frac{\rho}{\rho_{\oplus}}\right)^{-1}\left(\frac{M}{M_{\oplus}}\right).
\end{equation}
where $\rho$ denotes the density of the planetesimal material. We can now calculate the collision time scale for different populations. A promising case for creating a halo-like structure appears to be M2, which has high inclinations but is still located relatively close to the star. From table \ref{table:simulations} we find $\langle a\rangle=8.61$ AU, $\langle e\rangle=0.538$ and $i_{\rm max}=35.25^{\circ}$. For this specific configuration of comets, the collision time becomes
\begin{equation}
t_{\rm coll}^{(M2)} =3.7\times10^{5}\left(\frac{\eta}{1.41}\right)^{-1}\left(\frac{R}{\rm{km}}\right)\left(\frac{\rho}{0.5\rho_{\oplus}}\right)\left(\frac{M}{50M_{\oplus}}\right)^{-1}\rm \, yr.
\end{equation}

The collision time scale for the population in M2 is comparable to the migration time itself. This means a number of asteroids might suffer collisions during the migration itself, which we have not considered. The collision time is also an indication of how long the dust can survive in the system, as new dust is continuously formed during this period.

	\subsection{Dust Size Distribution}\label{sec:dsd}
When planetesimals collide at sufficiently high velocities, they will fragment and create smaller bodies. The distributions of these bodies has the shape of a power law
\begin{equation}
 f(m)\propto m^{-q},
\end{equation}
with $q=11/6$ \citep{dohnanyi68}. Assuming spherical particles we can rewrite this as a size distribution
\begin{equation}
 f(R) = f_{\rm R}R^{\gamma},
\end{equation}
with $\gamma=2-3q=-3.5$. The proportionality factor $f_{\rm R}$ depends on the material properties of the initial population and is calculated by \citet{dominik03} as 
\begin{equation}
 f_{\rm R} = N\sqrt{\frac{2m\sigma_{\rm coll}}{(4\pi/3)\rho\epsilon_{0}}}=NR^{5/2}\sqrt{\frac{8\pi}{\epsilon_{0}}},
\end{equation}
with
\begin{equation}
 \epsilon_{0}=-\frac{\pi \epsilon^{\gamma + 1}}{\gamma + 1},
\end{equation}
and
\begin{equation}
 \epsilon = \frac{v_{\rm coll}^2}{4S} - \sqrt{\frac{v_{\rm coll}^4}{16S^{2}}-\frac{v_{\rm coll}^2}{2S}}-1,
\end{equation}
where $S$ is the binding energy of the material. For asteroid type bodies $S=10^{7}\rm \,erg\,g^{-1}$ \citep{dominik03}.

    \subsection{Total Mass and Optical Depth}
The size distribution can be integrated to yield the total mass
\begin{equation}
 M_{\rm dust} = \int_{R_{\rm min}}^{R_{\rm max}} \frac{4\pi}{3}R^{3}\rho f(R)dR,
\end{equation}
in dust grains with radii between $R_{\rm min}$ and $R_{\rm max}$.

The smallest size, $R_{\rm min}$, is determined by comparing the gravitational force acting on the dust grain with the radiation pressure
\begin{equation}
\beta(R) = \frac{F_{\rm rad}}{F_{\rm grav}} = \frac{3L_{*}Q_{\rm pr}}{16\pi c GM_{*} R \rho}, 
\end{equation}
with $c$ the speed of light and $Q_{\rm pr}$ the radiation pressure coefficient
\begin{equation}
 Q_{\rm pr} \equiv Q_{\rm abs} + Q_{\rm scat}(1-\langle\cos \alpha \rangle),
\end{equation}
where $\alpha$ denotes the scattering angle. For isotropic forward scattering, $\langle \cos \alpha \rangle = 1$ \citep{burns79}. Particles with $\beta>1/2$ are ejected from the system.

Assuming $Q_{\rm pr}=1$, a dust grain with $\rho=2.76\rm \,g\,cm^{-3}$ orbiting a sun-like star reaches the critical value of $\beta=1/2$ for a radius of about $0.4\rm \,\mu m$. From here on on we will assume $10\rm \,\mu m$ to be the maximum size of a dust particle still contributing significantly to the optical depth.

Again considering M2, we have $v_{\rm coll}=14.4\rm \,km\,s^{-1}$. Together with $S=10^{7}\rm \,erg\,g^{-1}$ this gives $\epsilon_{0}=1.90\times10^{4}$. If we take the total planetesimal mass to be $50M_{\oplus}$, and the material density of the bodies to be $2.76\rm \,g\,cm^{-3}$, we find the total mass in dust grains between $0.4$ and $10\rm \,\mu m$ to be $M_{\rm dust}=8.7\times10^{-10}M_{\odot}$. This mass compares well to the mass in small dust grains quoted in section \ref{sec:halos}, but depends heavily on parameters such as the total planetesimal mass, and is only intended as an order of magnitude estimate.

The optical depth along a line of sight $dr$ can be written as
\begin{equation}
 d\tau = -\kappa \rho dr = n \sigma dz,
\end{equation}
with $\kappa$ the opacity and $n$ the number density of the particles responsible for the absorption. Assuming the dust is evenly distributed we can rewrite this as
\begin{equation}\label{eq:dtau}
 d\tau = \frac{\sigma_{\rm dust}}{V} dr, 
\end{equation}
where $V$ is the volume in which the dust is located, and $\sigma_{\rm dust}$ is the total combined cross-section for radiation of the dust grains
\begin{equation}
 \sigma_{\rm dust} = \int_{R_{\rm min}}^{R_{\rm max}} \pi R^{2} f(R)dR.
\end{equation}
Assuming the dust traces the planetesimal population, i.e., destructive collisions happen throughout $V$, we can integrate eq. \ref{eq:dtau} find
\begin{equation}
\tau^{(M2)}=9.6\times10^{-3}\left(\frac{R}{\rm{km}}\right)^{-0.5}\left(\frac{\rho}{2.76\rm{\,g\,cm^{-3}}}\right)^{-1}\left(\frac{M}{50M_{\oplus}}\right).
\end{equation}
The radial optical depth is indeed $<1$. Again, the derived optical depth depends a lot on the initial parameters. Since we have assumed the dust density is constant in the volume $V$, the radial optical depth is almost independent of the angle away from the midplane. 

The optical depth could be increased by making the planetesimal population more massive, or changing their size. This will however also increase $M_{\rm dust}$. 

A better way to increase $\tau$ might be to confine the volume in which the dust is located. This can be done in two ways; either moving the entire planetesimal population closer to the star, or by realizing the planetesimals have pretty eccentric orbits and most destructive collisions might therefore happen near periapsis. \citet{wyatt10} showed, using a Monte-Carlo simulation, that a population of bodies with identical semi-major axes and eccentricities, but random mean longitudes, arguments of pericentre and longitudes of ascending nodes, 90\% of destructive collisions happen within a radius $r/a = (1-e^2)/(1-0.72e)$, which might be a hint in this direction, even though the bodies in the populations in table \ref{table:simulations} are distributed both in $a$ and in $e$.

\section{Discussion}\label{sec:discussion}
Comparing the SEDs of several YSOs to radiative transfer models and near-IR interferometric visibilities appears to indicate the presence of an optically thin dust structure with a large scale height (section \ref{sec:halos}). We have argued that an inwardly migrating planet could explain the observed structure. The scenario is that an inwardly migrating planet collects planetesimals in mean motion resonances, raises their eccentricities, and subsequently scatters them to orbits with greater inclinations (fig. \ref{fig:sketch}). Over time, the planetesimals collide and form small dust in a collisional cascade.

	\subsection{The role of eccentricity pumping}
An important result of this study is that scattering planetesimals efficiently into highly-inclined orbits requires the relative velocity between the planet and the planetesimal to be high, of order of the local Keplerian speed. This is the reason why in our setup the sequence of eccentricity pumping in resonances followed by dynamical scattering is the most successful strategy.

The resonance capture probability of a migrating planet is a function of migration rate, planet mass, and distance to the star, and increases for more massive planets \citep{mustill10}. This puts constraints on the planet mass and migration rate. The process of eccentricity-pumping is well-defined by \citet{malhotra95,wyatt03} and eq. \ref{eq:epump} can be used to estimate the distance over which a particle has to be dragged in a particular MMR for its orbit to reach a certain eccentricity. However, if a planetesimal gets stuck in a resonance that is too far from the planet, it is likely to be accreted by the central star after its eccentricity approaches unity. This fate could possibly be averted by introducing gas drag, or a second, stationary planet close to the central star. In this case, the migrating planet would excite and push the planetesimals inwards, while the stationary planet would cause the scattering events. The mass of this second planet should not be too high, as this would lead to the ejection of the excited planetesimals. Systems with Neptune-mass planets within an AU of their parent stars are not uncommon \citep[e.g.][]{bouchy09}, and adding such a planet to the simulations presented in this work could well lead to even higher planetesimal inclinations.

\citet{yu01} study resonant capture by a Jupiter-mass planet migrating inward using analytic arguments and three-body integrations. They find that in typical systems, the resonance capture and subsequent eccentricity-pumping will most likely result in accretion of the test particle by the central star. This fate can be avoided by having a close encounter with the migrating planet, which eventually leads to an escape from the system. In the study presented here, systems with very massive planets show very similar behaviour.
	
	\subsection{Migration}
Our numerical simulations show that the basic scenario does work well for a certain range in planet mass and migration rate. The migration rates used are consistent with migration time scales for type I, type II, and migration through a planetesimal disk. However, the migration itself is not done self-consistently in our simulations. Because of this we have not taken into account the effect of the driver for the migration on the planetesimal population.

One possible driver is gas. A planet in a gaseous disk can exchange energy and angular momentum with the gas. If the planet is unable to significantly alter the disk structure it is referred to as type I. In this scenario the exchange of angular momentum can be written as the sum of the torques exerted by the Lindblad resonances of the planet on the disk \citep{goldreich79,goldreich80}. The direction of the migration depends on the size of the interior and exterior torques and is found to be inward for an isothermal disk \citep{ward97}. \citet{paardekooper06} simulated non-isothermal disks and showed the direction of migration depends on the ability of the disk to radiate away generated heat. \citet{tanaka02} calculated the sizes the torques for isothermal disks and derived migration time scales of the order of 1 Myr. The migration time scale is inversely proportional to planet mass, massive planets migrate faster \citep{armitage10}.

A slightly more massive planet will be able to open up a gap in the disk. In this case the co-rotating torque disappears. The migration rate for type II migration is equal to the gas-inflow velocity and found to be independent of planet mass \citep{armitage10}. Type II migration is generally faster than type I and at 5 AU a typical disk results in a migration time scale of $10^5$ yr.

Gas is not necessarily needed for migration though. A planet embedded in a planetesimal belt can exchange angular momentum with these bodies during scattering events. If the planet scatters bodies inward and outward at an equal rate, the net change in the angular momentum of the planet will be zero. If however, the planet usually scatters bodies to larger orbits, it will slowly loose angular momentum and migrate towards the star \citep{ida00}. For the change in angular momentum to have a significant effect on the planet's orbital radius, the mass of the planet has to be lower or equal to the total mass of scattered bodies. Also, for the migration not to halt the planet has to encounter fresh planetesimals continuously. \citet{ida00} calculate the migration rate to be
\begin{equation}
\frac{da}{dt}=4\frac{M_{\rm disk}}{M_{\odot}}\frac{a_{\rm pl}}{P_{\rm pl}},
\end{equation}
with the disk mass equal to $M_{\rm disk}=\pi \Sigma a_{\rm pl}^2$. \citet{kirsh09} study the orbital evolution of an isolated planet in a planetesimal belt and confirm this result, adding that for typical planetesimal surface density profiles the direction of migration will generally be inward.

It is stressed that the planetesimals that are having their eccentricities raised in an interior resonance are \emph{not} the ones responsible for the migration. Only when these bodies reach very eccentric orbits and suffer close encounters with the planet are they able to contribute to the planetary migration.

Aside from planet-disk interactions, the semi-major axis of a planet can also be changed significantly as a result of interactions with other planets. \citet{raymond11} numerically investigated systems with an initially unstable configuration of giant planets within a debris disk. It was found that the dynamical instability could trigger a collisional cascade.

	\subsection{The influence of gas in the disk}
While planetesimal scattering alone can make a planet migrate, the preferred scenario may be to have the planet migrate in the phase when the disk is still gas-rich, while our simulations with forced migration have ignored the effects of gas on the planetesimals. If we attribute the planet's radial motion to type I, type II, or planetesimal-driven migration in a gas-rich disk, we need to consider the effects of the gas during the migration and scattering phases. The presence of gas will cause aerodynamic gas drag. This drag will have two main implications; it will dampen the eccentricity and inclination of a planetesimal, and, it will have an effect of the probability of capturing a planetesimal in an MMR. Since the capturing of planetesimals followed by the eccentricity-pumping plays a vital role in this work, the effects of gas on these two mechanisms have to be addressed.

The aerodynamic gas drag time scale is given by \citet{adachi76}
\begin{equation}\label{eq:taero}
 \tau_{\rm aero} = \frac{8\rho R}{3 C_{D}\rho_{\rm gas}v_{\rm K}},
\end{equation}
with $\rho_{\rm gas}$ and $\rho$ the density of the gas and planetesimal material respectively. The drag coefficient $C_{D}$ is of order unity, and a non-linear function of the planetesimal's radius $R$ and relative velocity to the gas. From the equation it is clear that gas drag will predominantly effect smaller bodies at relatively small distances to the star.

\citet{capobianco11} study planetesimal driven migration in a gas disk and find the eccentricity of a planetesimal is dampened by a factor of order itself on a time scale of 
\begin{equation}\label{eq:taue}
\tau_{e} = \frac{2.3\times10^{2}\rm{yr}}{f(a,e)}\left(\frac{1.0}{f_{g}}\right)\left(\frac{R}{\rm{km}}\right)\left(\frac{a}{\rm{AU}}\right)^{13/4},
\end{equation}
where $f_{g}=1$ corresponds to a MMSN-disk and $f(a,e)$ is given by
\begin{equation}\label{eq:fae}
f(a,e) = \sqrt{\eta_{0}^{2}\left(\frac{a}{\rm{AU}}\right) + 7.5\times10^{-5}\left(\frac{M_{\rm pl}}{M_{\oplus}}\right)^{2/3} \left(\frac{e}{\chi}\right)^2},
\end{equation}
with $\chi = R_{\rm H}/a_{\rm pl}$ and $\eta_{0} = 1.95\times10^{-3}\rm \,AU^{-1/2}$. The damping time scale is smallest for small bodies since they couple to the gas better. For a larger planetesimal of 10 km with a semi-major axis of 10 AU, the damping time scale is $\sim10^{6-7}$ yr and the system approaches the gas-free case \citep{capobianco11}. It appears that the eccentricity-damping plays a small role for planetesimals of sizes $>1-10$ km. For smaller planetesimals or higher gas densities, gas drag will play an important role. From eqs. \ref{eq:taue} and \ref{eq:fae} it is obvious gas drag will affect eccentric orbits more than circular ones. In such a regime, eccentricity-damping could therefore protect planetesimals against accretion by the central star. This would allow for them to be scattered by the migrating planet, even if they are initially captured in a resonance far from the planet.

The coupling of planetesimals to the gas has an effect on the probability of resonance capture as well. \citet{capobianco11} define a radius $R_{\rm trap}$ for which half of the planetesimals are captured. For a MMSN disk this radius corresponds to 
\begin{equation}
R_{\rm trap} \simeq 9.2 \left(\frac{C_D}{0.5}\right)\left(\frac{\rho}{0.5\,\rm{g\,cm^{-3}}}\right)^{-1}\left(\frac{M_{\rm p}}{M_{\oplus}}\right)^{-0.73}\left(\frac{a}{\rm{AU}}\right)^{-11/4} \rm \,km,
\end{equation}
with $C_D$ the drag coefficient of order unity. For a Neptune-mass planet at 15 AU, $R_{\rm drag}\sim0.01$ km. Assuming the planetesimals are of km-size, the effect of gas on the capture probability can thus be neglected. 

\citet{fogg05} have studied the effect of a giant planet migrating in on the process of terrestrial planet formation, using N-body simulations. The inward migration is implemented artificially with a prescription that is chosen to match type II migration. The disk is assumed to be gas-rich. Consequently, the gas drag causes radial drift of planetesimals, and a damping of eccentricities and inclinations. \citeauthor{fogg05} find that the inward migration causes the planet to capture small bodies in its inner resonances, after which the orbits of these bodies are excited. The excitation of the orbits, followed by direct scattering by the planet, causes the formation of an exterior scattered disk of planetesimals, which looks very similar to the resulting populations in figure \ref{fig:all_BW}. Depending on the initial disk properties, the exterior disk is found to contain 30-70\% of the inner disk material. The gas damping is found to play an important role in the inner regions of the disk. For increasingly mature disks, \citeauthor{fogg05} find a wider spread in the orbital parameters of the scattered disk, caused by the increasingly smaller effect of gas damping.

In a somewhat similar study, with the focus of forming Earth-like planets in the habitable zone, is conducted by \citet{mandell07}. They find similar results for an inwardly migrating Jupiter-mass planet as \citet{fogg05}, including a scattered exterior disk. They also show that a second, stationary giant planet outside of the migrating one, will remove this scattered disk by ejecting the bodies from the system. A smaller or vanishing gas drag leads to more material ending up in the outer disk. The spread in orbital parameters of the bodies in the scattered disk also increases when less gas is present.

The extended scattered disks in these bodies show similar characteristics as the populations found in table \ref{table:simulations}, when gas does not play a big role. However, these studies focus on giant planet migration, in relatively young disks, thus focussing on low-velocity collisions between the planetesimals that allow them to grow and form planet embryos and Earth-like planets. In the study presented in this paper, we are instead focussing on less massive planets, and interested in the orbital parameters of planetesimals in the scattered disk.

After the migration has taken place, the planetesimals collide on a certain time scale to form dust. The dust distribution can be calculated under some approximations and the total mass and radial optical depth are found to be consistent with the observed haloes for certain values of variable parameters such as the total mass in the disk, the initial size of the planetesimals, and the planetesimal material density. The collisional model used in section \ref{sec:dust}, and the estimates of the optical depth and total mass are very simple and have their limitations. For instance, we assume the dust to trace the comet population, use simple geometrical estimates to define the volume the planetesimals occupy, and neglect the effect of any gas that might be present. Because of the nature of the discussed systems, including the spread in semi-major axes, eccentricities and inclinations, it is difficult to avoid these simplifications without having to make a detailed collisional and radiative transfer model, which is beyond the scope of this paper. If gas is present in the inner regions of the system, it will have a significant effect on the small dust grains since they couple to the gas on short time scales. Gas which is confined to the disk might rapidly remove small dust grains from highly-inclined orbits and reduce the scale height of the optically thin structure. Since we are dealing with pre-transitional disks, it is very well possible the inner regions of the system do not contain any gas at all.

\section{Conclusions}\label{sec:conclusions}
The goal of this work was to see if an inwardly migrating planet could be the cause of the observed optically thin halo-like structures. The main findings are: 
\begin{itemize}
\item An inwardly migrating planet is capable of capturing and transporting material inward in mean motion resonances. The probability of capture depends on planet mass and migration rate. Larger planet masses and lower migration rates increase the capture probability. The presence of gas on the capture probability can be ignored for km-sized planetesimals.
\item When a trapped planetesimal is pushed inwards, its eccentricity will rise rapidly and in agreement with equation \ref{eq:epump}. During this phase the inclination does not reach values greater than a couple of degrees. The growing eccentricity will eventually put the planetesimal on a planet-crossing orbit while increasing the relative velocity between the two.
\item Simulations show that the inward migration of a planet can lead to a population of planetesimals on highly-inclined orbits. The average inclination of the remaining population increases for higher planet masses and lower migration rates. Resonance capture and subsequent eccentricity pumping plays an important role in this outcome.
\item The high planetesimal inclinations in these systems result from direct scattering events, and reach significantly higher values if the planetesimal orbits are excited prior to this event.
\item In single-planet systems where planetesimals get stuck in an MMR far from the migrating planet, the vast majority of planetesimals is lost to the central object. This fate could perhaps be averted by introducing a second, stationary planet close to the central star. This planet should not be very massive, as close encounters would then lead to ejection of the excited planetesimals.
\item The resulting population of planetesimals on highly-inclined orbits have collisional time scales of the order of $10^{5-6}$ yr depending on the properties of the individual planetesimals and their combined mass. Since the relative velocities between the planetesimals are high ($\sim10\rm \,km\,s^{-1}$) the collisions will be destructive and will start a collisional cascade that will grind the bodies down to dust grains. The total mass and optical depth of this dust can be estimated and found to be of the same order as the total mass and optical depth observed in various systems.
\end{itemize}

Future studies should focus on self-consistent migration and a more careful description of the collisional cascade and optical depth determination to see if this mechanism can work and constrain the parameter-space further. With these future studies, it might be possible to use the optically thin structures as signposts for planet formation and migration. The presence of an optically thin halo-like structure could be induced from an SED, without having to resolve the system. The properties of the halo could then put constraints on the planet mass, migration rate, and distance it has travelled.

\begin{acknowledgements}
The authors would like to thank M. Wyatt for useful discussions, and the referee, P. J. Armitage, whose comments have led to a significant improvement of the manuscript.
\end{acknowledgements}

\bibliographystyle{aa}
\bibliography{refs}

\end{document}